\documentclass{article}

\usepackage{arxiv}

\usepackage[utf8]{inputenc} % allow utf-8 input
\usepackage[T1]{fontenc}    % use 8-bit T1 fonts
\usepackage{hyperref}       % hyperlinks
\hypersetup{
    colorlinks=true,
    allcolors=[rgb]{0.2,0.3,0.5},
}
\usepackage{url}            % simple URL typesetting
\usepackage{booktabs}       % professional-quality tables
\usepackage{amsfonts}       % blackboard math symbols
\usepackage{nicefrac}       % compact symbols for 1/2, etc.
\usepackage{microtype}      % microtypography
\usepackage{lipsum}		% Can be removed after putting your text content
\usepackage{graphicx}
\usepackage{natbib}
\usepackage{doi}
\usepackage{enumitem}\setlist{leftmargin=14pt}
\usepackage{pifont}

\title{Attribution-by-design: Ensuring Inference-Time Provenance in Generative Music Systems}

\author{
    \vspace{2.5mm}
    Fabio Morreale, Wiebke Hutiri, Joan Serrà, Alice Xiang, \& Yuki Mitsufuji \\
    \vspace{1.5mm}
    Sony AI \\
    {\small\texttt{\{fabio.morreale,wiebke.hutiri,joan.serra,alice.xiang,yuhki.mitsufuji\}@sony.com}}
}

\date{}

\renewcommand{\headeright}{}
\renewcommand{\undertitle}{}
\renewcommand{\shorttitle}{Attribution-by-design}

\hypersetup{
pdftitle={Attribution-by-design},
pdfsubject={Music attribution for generated contents},
pdfauthor={Morreale et al., 2025},
pdfkeywords={Data attribution, generative modelling, music, generation, compensation, transparency},
}

\begin{document}
\maketitle

\begin{abstract}
The rise of AI-generated music is diluting royalty pools and revealing structural flaws in existing remuneration frameworks, challenging the well-established artist compensation systems in the music industry. Existing compensation solutions, such as piecemeal licensing agreements, lack scalability and technical rigour, while current data attribution mechanisms provide only uncertain estimates and are rarely implemented in practice. This paper introduces a framework for a generative music infrastructure centred on direct attribution, transparent royalty distribution, and granular control for artists and rights' holders. We distinguish ontologically between the training set and the inference set, which allows us to propose two complementary forms of attribution: training-time attribution and inference-time attribution. We here favour inference-time attribution, as it enables direct, verifiable compensation whenever an artist’s catalogue is used to condition a generated output. Besides, users benefit from the ability to condition generations on specific songs and receive transparent information about attribution and permitted usage. Our approach offers an ethical and practical solution to the pressing need for robust compensation mechanisms in the era of AI-generated music, ensuring that provenance and fairness are embedded at the core of generative systems.
\end{abstract}

% keywords can be removed

\section{Introduction}

Users typically interact with modern generative AI systems by describing in natural language the characteristics of the media they want to generate. This textual prompt - e.g.,~\textit{``Create a musical piece in the style of Mozart for a graduation ceremony''} - is fed into the generative model to condition the generation. While this interaction paradigm has yielded satisfactory outputs, it also has significant ethical and business shortcomings \citep{sturm2019artificial,drott2021copyright,morreale2021does,clancy2023artificial}. In an ideal scenario (one that is ethical, fair, and economically viable), creators would receive compensation every time their work meaningfully influences the generation of a new piece. However, in practice, the dominant paradigm of generative AI relies on the large-scale scraping of cultural materials that are accumulated into massive and ill-documented datasets \citep{gebru2021datasheetsdatasets,Scheuerman_2021}. These practices typically proceed without seeking consent from creators, without providing transparency about which works are included, and without establishing mechanisms for remuneration~\citep{morreale2024unwitting}. As a result, most creators are neither aware nor compensated for the use of their art in generative models~\citep{morreale2023data,10.1145/3706598.3713799}. Even when such awareness exists, current generative systems lack any reliable means of identifying which human-created art has contributed to the generation of new outputs, and to what extent. The consequences are both ethical, in the form of unconsented labour, and financial, in the form of unpaid labour~\citep{morreale2024unwitting}. Within this regime, generative systems function primarily as competitors to human artists, offering them little to no tangible advantage.

The problem of attribution
% in practical terms, thus becomes one of distributing royalties to creators whose art has been used to produce, in one way or another, some generated content. This problem 
is particularly salient in the context of music. The music industry indeed already operates on a deeply entrenched system of rights management and royalty distribution. Unlike other domains, music offers a well-established legal and institutional infrastructure: composers, lyricists, performers, and other rights holders are routinely compensated for uses of their work in broadcasting, streaming, sampling, and public performance. This pre-existing royalty logic makes it untenable to treat AI training datasets as if they fall outside of this framework. Thus, in the remainder of the article, we will specifically focus on music applications. However, our proposed solution is not music-specific and could be adopted \textit{mutatis mutandis} in other domains.

Existing approaches to attribution in generative modelling remain limited both epistemically and technically. Most current methods attempt to infer influence retrospectively, either by measuring similarity between generated and training data (corroborative approaches) or by estimating the contribution of individual training samples through model introspection (predictive approaches). Both strategies rest on problematic assumptions: similarity does not imply causation, and model internals are not necessarily interpretable in human terms. As a result, these methods can only provide probabilistic or speculative notions of provenance, which are inadequate for establishing transparent or enforceable compensation mechanisms. Such uncertainty renders post-hoc attribution both unreliable and impractical, thus warranting the explorations of new paradigms.

In this paper, we introduce one such paradigm. Attribution-by-design is a framework that directly embeds attribution into the architecture of generative music systems. This paradigm reframes attribution as a constitutive property of generative modelling rather than a retrospective operation, aligning epistemic transparency with established norms of  artistic authorship and legal accountability. In doing so, it proposes a way to re-integrate value recognition into the technical fabric of AI music generation. Our main contributions are threefold. First, we provide an ontological distinction between training and inference datasets and between training-time and inference-time attribution. Second, we elaborate how inference-time attribution can operate within this framework, defining the conceptual and methodological principles that make attribution verifiable and ethically grounded. Third, we outline the sociotechnical infrastructure required to implement this framework, encompassing user interaction, verification, generation, and compensation. 

The remainder of the paper is structured as follows. Section 2 reviews existing approaches to data attribution in generative modelling. Section 3 introduces our theoretical grounding and the attribution-by-design framework. Sections 4 to 6 detail its implementation through user interaction, verification and compensation mechanisms, and the generation process. Section 7 discusses limitations and future research directions, and Section 8 offers concluding discussions.

\section{Background: Attribution in Generative Modelling}

The challenge of identifying which training data has influenced a specific output of a machine learning model is commonly referred to as the \textbf{data attribution} problem~\citep{Hammoudeh_2024,ilyas_data_2024}. For generative modelling, existing approaches can be broadly grouped into two families: corroborative (or model-agnostic) and predictive (or model-based). \textit{Corroborative} approaches assess attribution without requiring access to the internals of the generative model, and generally compute similarities between generated outputs and training data. A common approach is to estimate similarities between latent space representations of the data, using multi-purpose~\citep{caron_emerging_2021,huang_mulan_2022,li_mert_2024} or ad-hoc~\citep{lu_deep_2017,serra_supervised_2025} encoders known to reflect, to some extent, certain notions of perceptual or musical similarity.\footnote{Another variant would leverage watermarking techniques for corroborative attribution~\citep{asnani2024promark}, exploiting the notion of watermark ``radioactivity''~\citep{sablayrolles2020radioactivedatatracingtraining}. However, as recent research shows strong limitations in audio watermarking~\citep{wen_sok_2025,oreilly2025deepaudiowatermarksshallow,ozer_comprehensive_2025}, we leave aside the discussion of such a variant.} In the music domain, corroborative pipelines have been explored to assess the potential influence of training material in generated outputs~\citep{barnett_exploring_2024,batlle-roca_towards_2024}. \textit{Predictive} approaches, by contrast, exploit knowledge of the specific generative model producing the generations under study, and leverage its structure and internals to derive an influence score between generated and training data. Common approaches revolve around analysing or comparing the impact of individual data points on the model's weights~\citep{wang_data_2024,ko_mirrored_2024} or its gradients~\citep{zheng_intriguing_2024,lin_diffusion_2024}. To date, in the music domain, there only exist a handful of predictive methods~\citep{deng_computational_2024,deng2024efficientensemblesimprovetraining,choi_large-scale_2025}. 

A critical examination of these attribution approaches reveals that they both involve unexamined assumptions. Corroborative methods presuppose a causal link between similarity and influence: they assume that training data points most similar to a given output must also have been the most causally responsible for producing it. However, this assumption is untenable~\citep{pearl2009causality}. Similarity measures quantify resemblance according to a predefined metric, but they do not (and cannot) guarantee that a specific data point actually shaped the model parameters that generated the output. Thus, higher similarity between an output $O$ and an input $I_a$ (compared to $I_b$) does not logically entail that $I_a$ had greater causal influence on the generation of $O$. Predictive approaches assume that the reasoning of a generative model is human-intelligible, or at least traceable, and that carefully-designed methods can thus identify the source of attribution. Yet, human and machine abstraction might be incommensurable, i.e.,~they might not be directly accessible to the other~\citep{fazi_beyond_2021}. Attempts at predictive attribution, therefore, risk imposing human interpretative reasoning on processes that are fundamentally inaccessible~\citep{burrell_society_2021,alvarado_ai_2023}. For these reasons, predictive approaches represent, at best, a biased notion of attribution or influence. Overall, %for both corroborative and predictive approaches, 
there can be no certainty that a specific training data has actually been used for a specific generation. 
%As a consequence, it is impossible to guarantee that a user’s wish (to create a song in the style of Mozart) has been respected. 
This lack of certainty has consequences for all involved parties: artists whose works have contributed to model training cannot ascertain whether they have been appropriately compensated (were such mechanisms in place), while users are unable to verify whether their payments ultimately reach the creators whose materials influenced the generated output.

\section{Attribution-by-design} 

This section introduces the attribution-by-design framework, outlining its conceptual foundations and describing how attribution can be embedded directly into the architecture and operation of generative music systems.

\subsection{Training vs.~Inference Dataset and Attribution}

\begin{figure}
    \centering
    \includegraphics[width=0.6\linewidth]{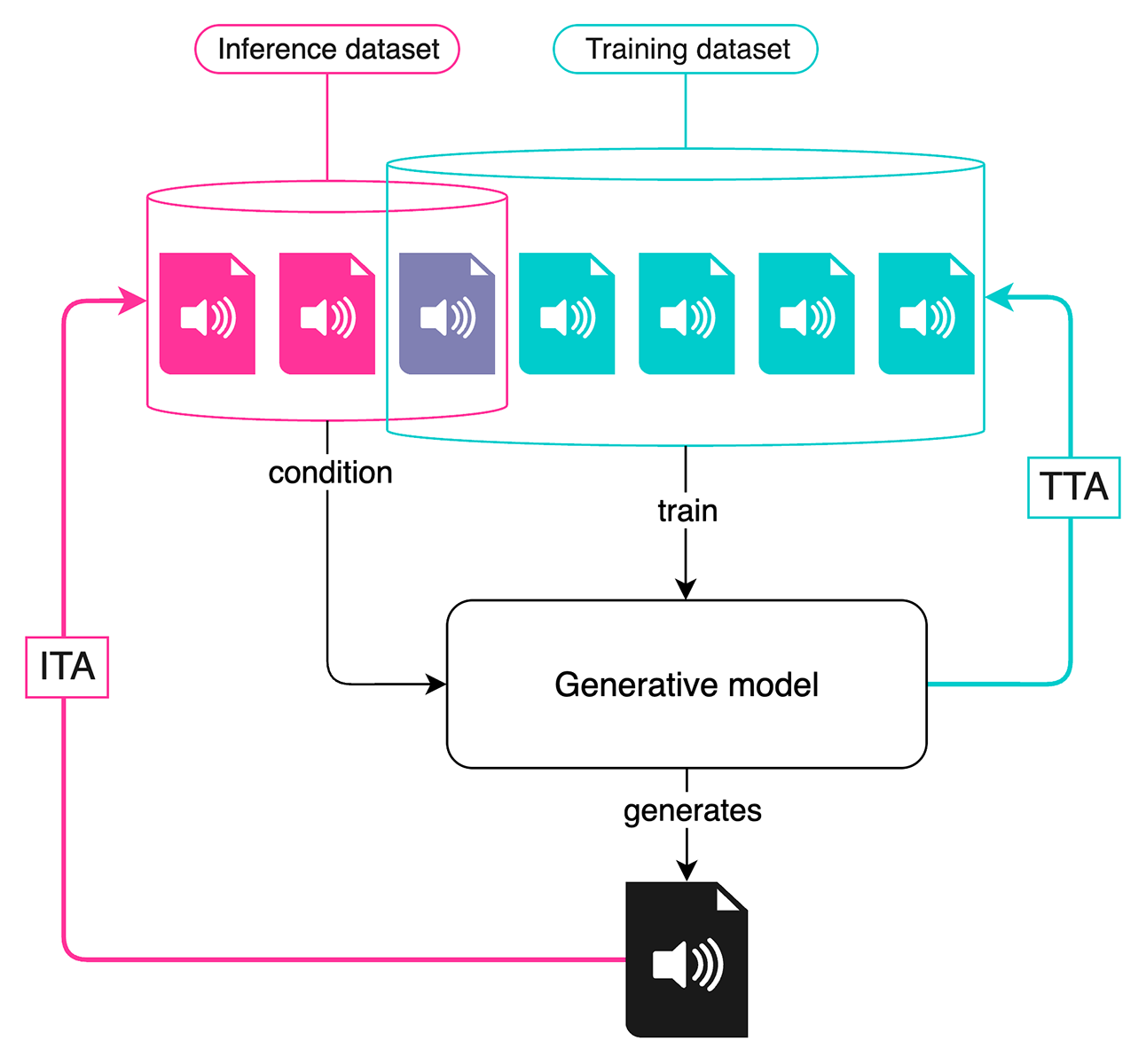}
    \caption{Two stages of attribution (TTA vs.~ITA), and separation between inference and training datasets.}
    \label{fig:2steps}
\end{figure}

Before outlining the proposed approach, we articulate two innovative ontological propositions that ground our work. The first proposition is an explicit separation of the training dataset from the inference dataset in generative modelling (Fig.~\ref{fig:2steps}). The \textbf{training dataset} is the collection of data from which the model learns. During training, these data are used to estimate the model’s parameters and to structure its internal representations. In other words, the training dataset defines what the model is able to encode and generalise from. The \textbf{inference dataset}, by contrast, refers to the collection of data available to the user at generation time. Such data is not used to shape the model parameters, but instead can be viewed as a catalogue that a user can browse to find the reference material that will shape a particular output. The two sets may overlap entirely, partially, or not at all. Nevertheless, they remain conceptually distinct: the training dataset pertains to the constitution of the model while the inference dataset pertains to its operation. Establishing this distinction is a necessary step for clarifying the second proposition.

The second proposition is a split between different kinds of attribution in generative modelling (Fig.~\ref{fig:2steps}). We term these \textbf{training-time attribution} (TTA) and \textbf{inference-time attribution} (ITA). While existing attribution techniques conflate these two levels, thereby blurring their ontological difference, we argue that they are, in fact, ontologically distinct.
TTA captures the contribution of a given \textbf{training dataset} point to the \textit{formation of the model itself}; that is, how much of what the model has internalised during training can be traced back to a particular data point from the training set. This is attribution at the level of \textit{model constitution}. ITA, by contrast, captures the contribution of an \textbf{inference dataset} point to the \textit{production of a specific output}; that is, how much of a given generation can be attributed to a data point from the inference set. This is attribution at the level of the \textit{event of generation}. These two forms of attribution are ontologically distinct, and identifying them demands different epistemological approaches. TTA concerns the data sedimented into the model’s parameters and representational space, whereas ITA concerns the mobilisation of that sediment in a concrete act of generation. Collapsing them under a single heading risks obscuring both their technical challenges and their ethical stakes. We therefore propose this distinction as a conceptual lens through which attribution methods should be framed.

Because TTA remains epistemologically and technically intractable, our proposed framework is grounded in ITA. ITA enables transparent and direct attribution, allowing the contribution of reference material to be explicitly traced from specific generated outputs. In a sense, our approach seeks to minimise the role of training data in the processing of inference material, a role that might not be attributable in principle. By doing so, we offer a pragmatic way to circumvent the limitations of TTA, while ensuring verifiable attribution at inference time.

\subsection{Framework} 

Our approach, which we term \textbf{attribution-by-design}, is an ethics-first, lifecycle-wide framework for generative models with ITA embedded at its core. In attribution-by-design, provenance is \textit{enforced}: the generative model is conditioned to generate in the style of specific song(s), which are directly selected by the users. Users directly select reference songs from an available catalogue (the inference set), and these songs are then provided as conditioning to a generative model to perform sample-based generation. Under this framework, the association between a generated output and reference songs is certain, rather than speculative, as there exists a direct connection between user inputs and the inference data source that contributed to the generated outputs. As the catalogue is accessible to the model owner, our proposed approach enables the design of transparent and enforceable mechanisms for compensating creators, based on contracts between contributing artists and the service provider. Ethical attribution is thus built into the core of the system at every step. 

Attribution-by-design has the following strategic advantages:
\begin{enumerate}
    \item It permits a contractual model in which generative model creators must engage with artists and right holders to license their work for training and/or inference, as well as fulfil other contractual obligations, such as attribution royalties and purpose limitations to specified uses. 
    \item It provides correct identification of contributing sources during the generation process, rather than attempting to reverse-engineer attribution after content has been created, and thus eliminates ambiguity about which specific inference data sources contribute to the generated output.
    \item It enables immediate and transparent attribution relationships, providing clarity for both users and rights holders about which sources contribute to specific generated outputs.
    \item It represents a step toward more sustainable creative ecosystems by ensuring that original creators receive appropriate compensation when their works contribute to generated music.
\end{enumerate}

\begin{figure}
    \centering
    \includegraphics[width=0.75\linewidth]{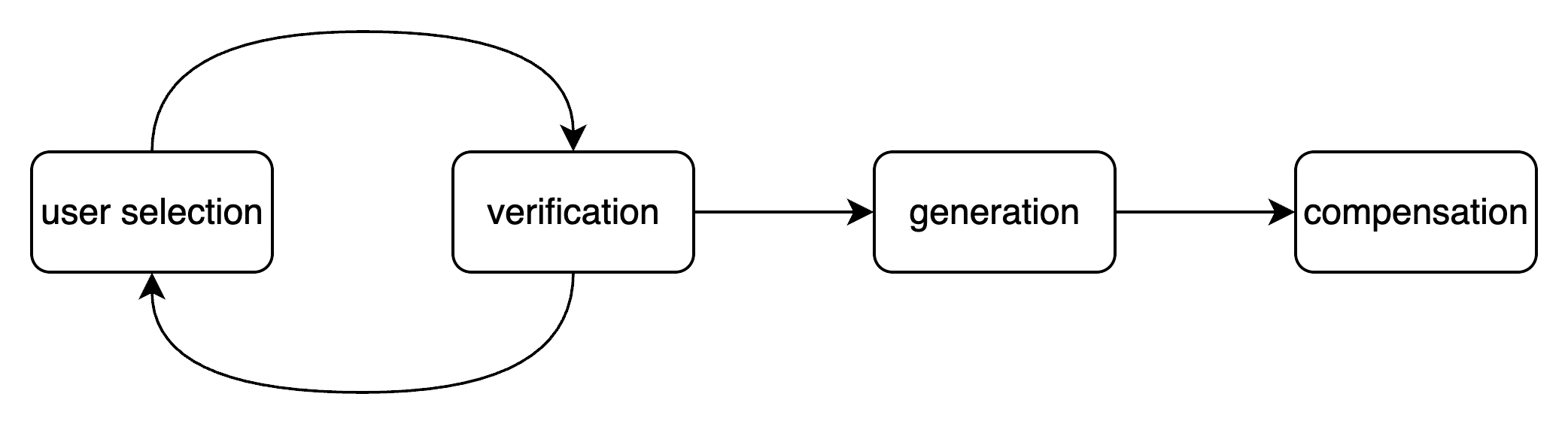}
    \caption{Attribution-by-design process.}
    \label{fig:flow}
\end{figure}

In the next sections, we present the attribution-by-design framework (Fig.~\ref{fig:flow}), which is structured as a four-step process 
% including:
% \begin{enumerate}
%     \item \textbf{User interaction ---} to receive information from the user about the specific song(s) they want to include as references. 
%     \item \textbf{Verification ---} to verify whether these songs meet the artists' condition for consent.
%     \item \textbf{Generation ---} to generate a new song conditioned on the referenced songs. 
%     \item \textbf{Disbursement ---} to compensate creators of referenced songs.
% \end{enumerate}
featuring \textbf{user interaction} (to receive information from the user about the specific songs they want to include as references), \textbf{verification} (to verify whether these songs meet the artists' condition for consent), \textbf{generation} (to generate a new song conditioned on the referenced songs), and \textbf{compensation} (to compensate creators of referenced songs).

\section{User Interaction}
A central challenge of attribution-by-design lies in collecting reference songs from the user. This problem entails a paradigm shift in the way attribution is conceived: it is no longer a matter of statistical similarity or other forms of guesswork, but one of sociotechnical infrastructure and user interaction. The user experience thus becomes particularly important, as it must allow users to navigate a vast space of millions of songs. We distinguish two main modes through which users can select reference songs: unmediated (direct) and mediated (indirect).

\subsection{Unmediated (Direct) Selection}

In this scenario, the user directly selects one or more reference songs from the catalogue to inspire the generation (or a specific part of it). As the songs to be rewarded are directly selected by the users, attribution is direct and unmediated. As a consequence, there is no need to rely on attribution models that run ad-hoc calculations. 

\subsubsection{Levels of Control}
\begin{figure}[h]
    \centering
    \begin{minipage}{0.48\linewidth}
        \centering
        \includegraphics[width=\linewidth]{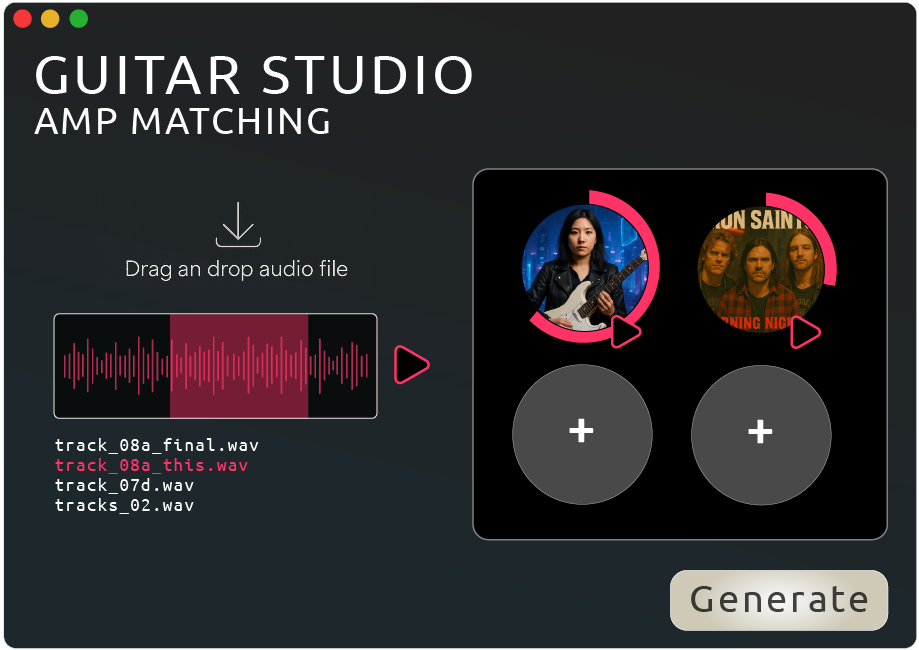}

    \end{minipage}
    \begin{minipage}{0.48\linewidth}
        \centering
        \includegraphics[width=\linewidth]{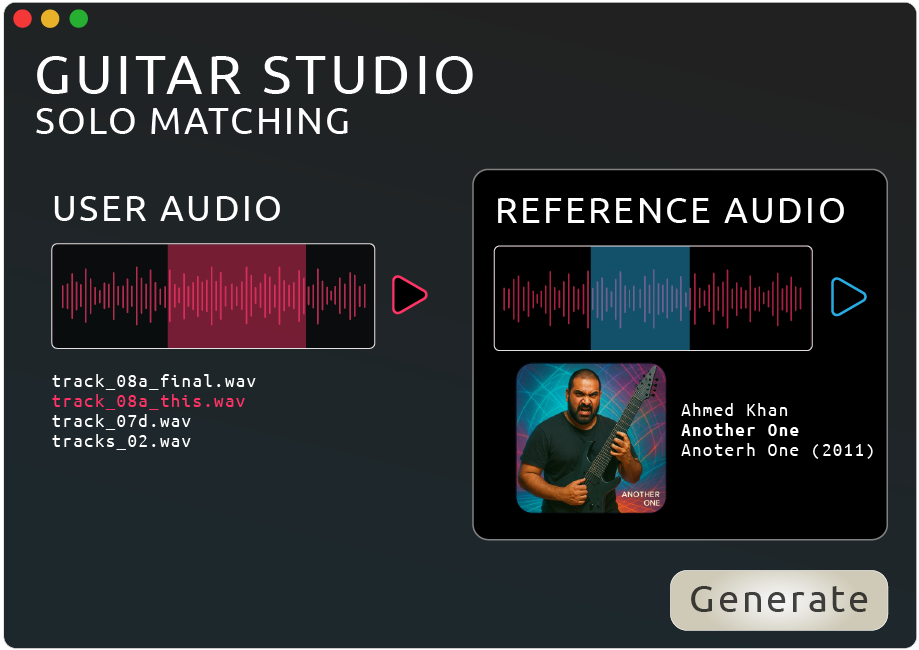}  
    \end{minipage}
    \caption{Two possible interaction possibilities for audio-level UIs.}
\label{fig:guitar}
\end{figure}

Different levels of control can be developed to cater to various user requirements and creative experiences: 
\begin{itemize}
\item \textbf{Song level ---} The user identifies one or more songs (reference material) from a catalogue (inference dataset) that they want the output to be inspired by (conditioned on). In Fig.~\ref{fig:selection} (left and center), the user has already selected two songs and is currently browsing the interface to add another reference song.

\begin{figure}[h]
    \centering
    \begin{minipage}{0.32\linewidth}
        \centering
        \includegraphics[width=\linewidth]{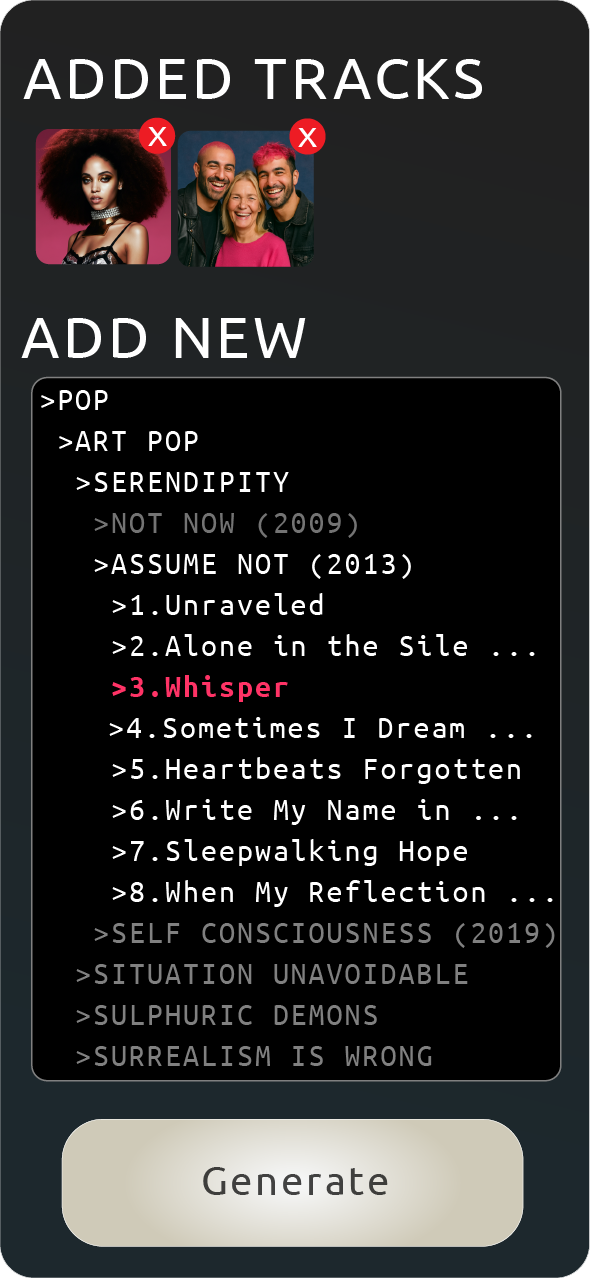}
    \end{minipage}
    \begin{minipage}{0.32\linewidth}
        \centering
        \includegraphics[width=\linewidth]{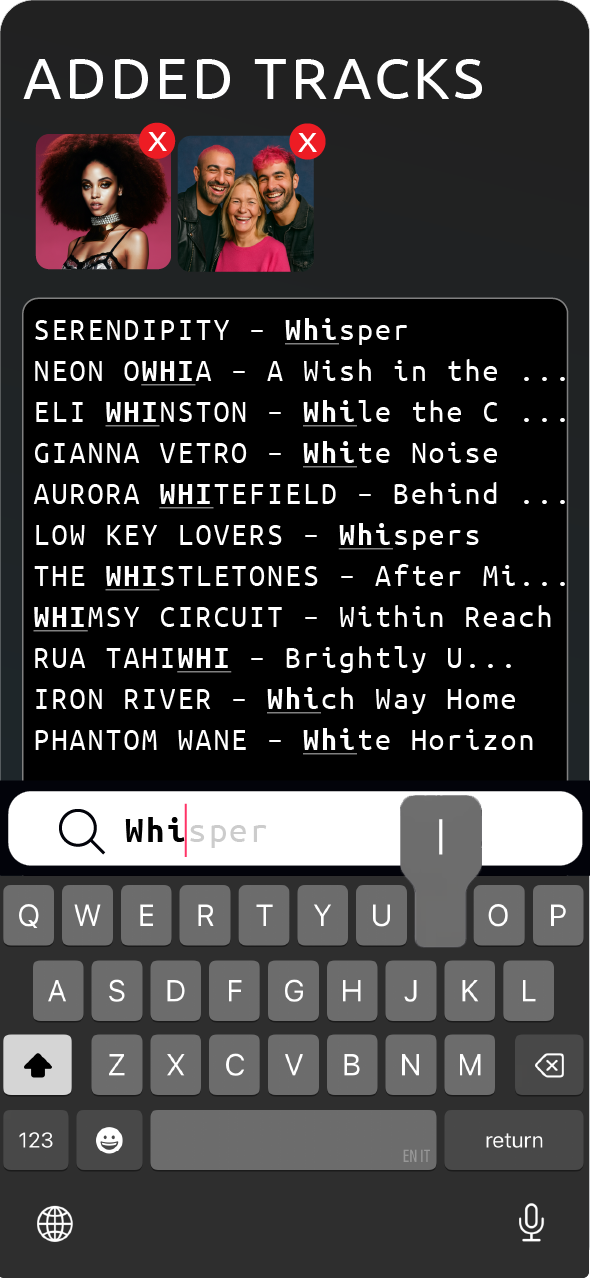}
    \end{minipage}
    \begin{minipage}{0.32\linewidth}
        \centering
        \includegraphics[width=\linewidth]{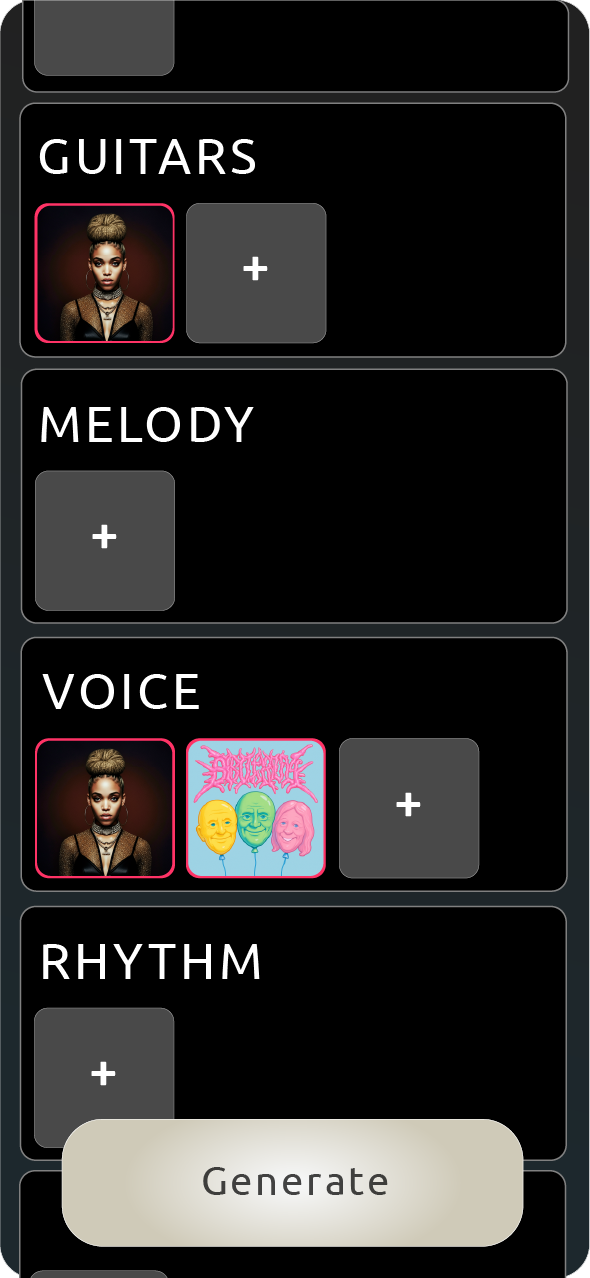}
    \end{minipage}
    \caption{(Left) Song-level selection using a hierarchical UI; (center) song-level selection using a search UI.; (right) parameter-level selection UI.}
    \label{fig:selection}
\end{figure}

\item \textbf{Parameter level ---} Rather than using all the song elements as a reference, the user might select one or more individual parameters from reference songs. Such parameters can encompass a range of stylistic, compositional, performative, and semantic elements. Thus, users would be enabled to specify particular aspects or characteristics of the songs they wish to influence the output. For instance, in Fig.~\ref{fig:selection} (right), the user selected one specific song to influence both the guitar and the voice, and a second song to also influence the voice. They did not specify any other parameters, such as melody or rhythm, which would then be generated unconditionally or selected procedurally.\footnote{We leave it open how to technically complement user queries. However, there are essentially two approaches to follow. On the one hand, the generative model can resort to training data and generate unconditionally for the non-specified parameters. This obviously puts more weight on TTA, which, as mentioned, is an unsolved topic and out of the scope of the current paper. On the other hand, additional songs can be procedurally selected in a random or automated way (potentially leveraging existing retrieval or recommendation engines; see below). This approach falls back to ITA and would follow the same attribution mechanism as the user-selected songs. It also allows for more transparency, as the system would now be able to tell which songs are used as references for the unspecified parameters.}

\item \textbf{Audio level ---}
The interaction could be more fine-grained, allowing the user to provide an audio track, with the system applying one or more attributable parameters - derived from the reference material or textual input - to the user’s media sample. Such a mechanism may enable users to enhance or modify their own music by applying specific characteristics from reference songs. For instance, in Fig.~\ref{fig:guitar} (left), the user uploaded a guitar stem and asked the system to perform amp matching (i.e.,~to analyse the tonal characteristics of a reference guitar recording and reproduce its amplifier response on the user’s track), specifying two reference songs with different weights.

\item \textbf{Temporal level ---} The interaction can be made even more granular by allowing the user to select a specific segment of the audio file they wish to modify, along with a corresponding segment of the reference song.  For example, in Fig.~\ref{fig:guitar} (right), the user prompts the system to edit a portion of an uploaded guitar solo so that it sounds more “in the style of” a chosen section of a solo from an existing song.

\end{itemize}

\subsubsection{Selection Mechanisms}

As the inference set can potentially contain millions of songs, the UI must be designed to allow users to discover and select appropriate songs from such a large inference catalogue. Several selection mechanisms can be developed: %, including: search-based, hierarchical structure, and recommendation engine.
\begin{itemize}

\item \textbf{Hierarchical structure ---} The user navigates a UI where songs are hierarchically organised in a tree-like structure. Fig.~\ref{fig:selection} (left) illustrates one such hierarchy, beginning with genres and subgenres, followed by artists and albums, and ending with individual songs. Other hierarchies can also be adopted, for example, based on release date (decade/year/month/song), geographical area (country/city/artist/song), or instrument-based (instrument family/instrument/performer/song/specific performance).

\item\textbf{Search-based ---} The user prompts the system to find a specific song using a standard text-based search bar. Possible inputs are song/album titles, artists' names, keywords, or other descriptions. An autocomplete function can be employed to facilitate and/or constrain the interaction (see Fig.~\ref{fig:selection}, centre).

\item \textbf{Recommendation engine ---} In this scenario, the system suggests potentially relevant songs or samples rather than requiring the user to actively search or browse through a hierarchy. The recommendation engine can follow typical approaches to music recommendation and thus be informed by different types of signals, such as the user’s past interactions, listening history, or explicit preferences expressed during the session. Notably, this approach involves ranking and filtering, and is therefore inherently mediated. %Particular care must be taken to address the biases that typically affect recommendation systems \citep{porcaro2021diversity}.

\end{itemize}

\subsection{Mediated (Indirect) Selection}\label{sec:mediated}
Generative AI companies explicitly target non-expert musicians and complete non-musicians, presenting their systems as democratising tools that \textit{unlock dormant creativity} and remove the need for musical training. While such invocations of democratisation often function as commercial rhetoric that serves business models while constraining musical practice~\citep{pram2025opening}, non-musicians, nevertheless, remain a plausible user base. Some users may lack not only formal musical knowledge but also the cultural reference points needed to select an appropriate reference song, which is required in the unmediated scenario. Yet, this category of user may more or less effectively communicate their creative intentions through textual descriptions. This model is already familiar from music streaming, where text queries or descriptive tags mediate access to playlists and music catalogues.  In our framework, these situations are captured by the mediated selection scenario. 

\subsubsection{Retrieval model}\label{sec:input_synth}
The user’s input is processed by a retrieval model that identifies and presents a subset of songs from the inference set deemed relevant to the prompt. The retrieval model can be implemented as a language-music model (LMMs), such as MuLan~\citep{huang_mulan_2022} or CLAP~\citep{elizalde2022claplearningaudioconcepts}, which maps audio and text into a joint embedding space, thereby enabling the retrieval of supposedly relevant songs from textual descriptions.

\begin{figure}[t]
    \centering
    \includegraphics[width=\linewidth]{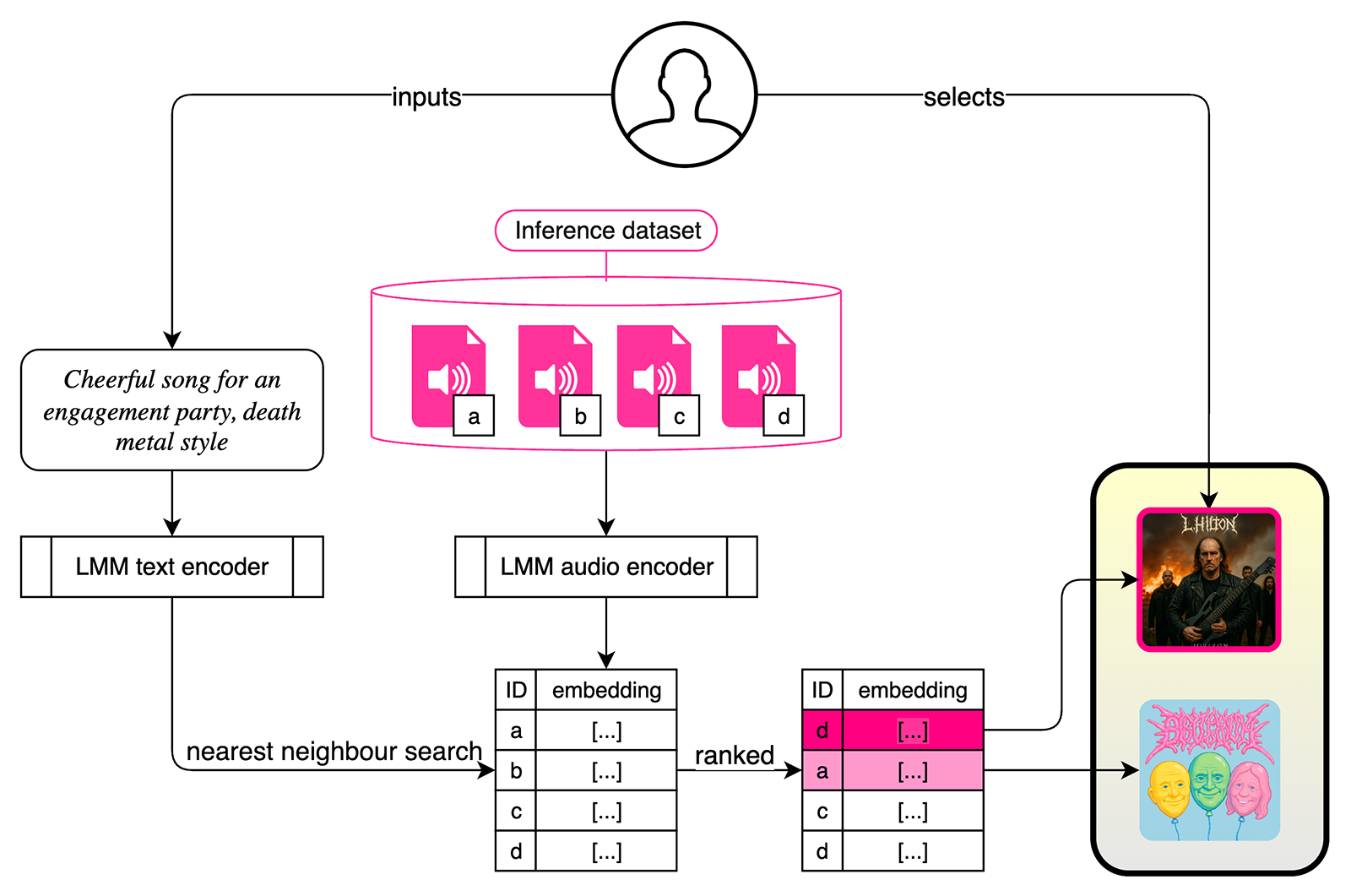}
    \caption{A diagram indicating how an LMM can retrieve relevant reference songs from the inference dataset.}
    \label{fig:clap}
\end{figure}

In practical terms, the retrieval model receives as input a prompt (see details in the next subsection) that describes what the user aims to generate. This prompt is encoded using the LMM’s text encoder (Fig. \ref{fig:clap}), and similarity scores are computed against the embeddings of all songs in the inference set. Songs with the highest similarity to the prompt's embedding are considered the most relevant and are then ranked as the candidates that most closely align with the user prompt. The first $K$ songs identified by the LMM are displayed to the user, who can preview them and request additional or alternative candidates if unsatisfied with the initial selection. This process is repeated until the user is satisfied with one or more songs and selects them as reference songs. Note that this approach still relies on similarity scores, but these scores are now solely used to navigate across candidate songs rather than to corroboratively determine attribution. Crucially, the user makes the final selection, ensuring that attribution remains \textit{by-design} and thus grounded in explicit user choices rather than in opaque algorithmic selection.

\subsubsection{User Prompts}
User prompts might be of two types: textual prompts and category selection. With \textit{textual prompts}, which are commonly found in current generative model applications, the user simply uses natural language to communicate to the system what they want to be generated. This approach offers the advantage of free-flowing, nuanced, and highly personalised requests, enabling users to express complex or subtle creative intentions without constraint. However, these requests need to be interpreted by a neural network (e.g., an LMM), whose outputs are shaped by epistemic uncertainty and may therefore fail to produce coherent or semantically grounded results.

An alternative way to collect prompts from the user is to allow them to select discrete options from predefined categories, allowing for more structured input. Category examples include: genre (and subgenres; found in metadata or curated from dedicated websites), instruments (identified via metadata; possibly integrated with information from dedicated databases), vocalist gender and age (e.g.,~male, female, non-binary, old, young, middle-age), listener's mood (e.g.,~happy, sad, energetic, nostalgic, angry, relaxed), situation (e.g.,~baby shower, wedding, funeral, birthday, graduation), or function (e.g.,~workout, relax, focus, study, sleep, yoga, running). Selecting from discrete labels or categories can improve retrieval accuracy and generation quality, as the system can more reliably match user intent with relevant reference material. Structured prompts reduce ambiguity and might facilitate faster user interaction. Furthermore, discrete options help standardise user input, making it easier to map requests to available catalogue metadata and to ensure compliance with rights management and attribution requirements.

\section{Verification and Compensation}

Before a selected reference song is released for use, the user input is subject to verification to ensure that the generation request respects artists' conditions for consent, distribution policies, and rights management requirements. This verification ensures that new songs can only be generated if artists and rights holders have explicitly opted in to make the selected reference songs available for generative inference, for a defined purpose. It further aims to align compensation mechanisms for generative inference with processes that are already the norm, e.g., in sampling, where artists receive financial benefit if their sampled music contributes to a new work. 
% Crucially, such alignment between purpose, rights, and remuneration becomes possible under the proposed Attribution-by-design paradigm, which embeds attribution and consent directly into the generative process itself.

\subsection{Nuanced Consent}
Instead of framing consent as an all-or-nothing decision imposed upon artists, attribution-by-design paves the way for fine-grained opt-in, checked against specific conditions and intended end use as elicited through the user interaction. Such nuanced opt-in verification breaks sharply from the status quo, which assumes consent by default, for any purpose, for all, forever. Here, nuanced consent is explicitly captured, for example, in a database where columns indicate whether a specific song or aspects of a song can be used, like in Table~\ref{tab:verification}. 
While defaults are set to an opt-out condition that prevents use, specific permissions can be directly obtained from and updated by artists, rights holders, or representatives who wish to partake in the generative AI ecosystem. Importantly, any updating mechanism must enable consent revocation, such that opt-in is not granted indefinitely, but only as long as rights holders deem desirable.

\begin{table}[t]
\centering
\begin{tabular}{c c c c c}\toprule
\textbf{Song ID} & \textbf{Model training} & \textbf{Song-level inference} & \textbf{Parameter-level inference} & \textbf{Audio-level inference} \\\midrule
17189 & \ding{55} & \ding{51} & \ding{51} & \ding{51} \\
17193 & \ding{55} & \ding{51} & \ding{55} & \ding{55} \\
17194 & \ding{55} & \ding{51} & \ding{51} & \ding{55} \\
17196 & \ding{55} & \ding{55} & \ding{55} & \ding{55} \\ \bottomrule
\\
\end{tabular}
    \caption{Attribution-by-design allows for hard-coding how a song can be used in the generative AI framework.}
    \label{tab:verification}
\end{table}

% \subsection{Validation}
% The selection of reference media is subject to validation. This validation is in place to ensure that such a selection aligns with the entries of the acceptable usage policies and rights management requirements. This approach limits user choices to pre-approved options, guaranteeing that creators' wishes are respected. 

\subsection{Distribution Policies}

In addition to nuanced consent, distribution policies can be specified for reference songs to account for the potential audience and revenue generated by these outputs. Table \ref{tab:intended_use} shows a few possible intended uses for generated songs. These include private use, non-commercial, and commercial distribution. If the intended use of a generated output is permitted by the distribution policies of the reference song, then this second verification step is cleared. 

Different usage scenarios and distribution channels can be priced distinctly, with commercial uses requiring higher compensation than private or preview uses. A tiered compensation system can make generated content more accessible for personal and educational uses while ensuring that commercial applications provide appropriate compensation to rights holders.  This purpose-based compensation model offers flexible remuneration and provides artists with complete control over the music that can be generated from their creations. 

\begin{table}[t]
\centering
\begin{tabular}{c c c c c}\toprule
\textbf{Song ID} & \textbf{Save for private use} & \textbf{Non-commercial distribution} & \textbf{Commercial distribution} \\\midrule
17189  & \ding{51} & \ding{51} & \ding{51} \\
17193  & \ding{55} & \ding{55} & \ding{55} \\
17194  & \ding{51} & \ding{51} & \ding{55} \\
17196  & \ding{51} & \ding{55} & \ding{55} \\  \bottomrule
\\
\end{tabular}
\caption{Rights holders have detailed control over the use of generated songs based on their catalogue.}
\label{tab:intended_use}
\end{table}

\subsection{Transparency as Precondition for Compensation}
As attribution-by-design precisely determines the contribution of reference songs \textit{prior to} generation, each generated output can be associated with rights holders, allowing them to receive compensation. However, merely receiving compensation is not a guarantee for \textit{fair} compensation. A trustworthy compensation mechanism requires transparency into how compensation is allocated. The outcomes of all attribution and verification steps must thus be logged, so that rights holders can validate and confirm attribution and subsequently compensation pay outs. When generated outputs are disseminated on digital platforms, further opportunities for compensation can arise, as is presently the case. To enable rights holders to share the benefit when generated outputs are disseminated, transparency mechanisms that embed sufficient metadata in generated outputs are necessary so that generated music can be traced on digital platforms.

\subsection{Recommendation and Discovery as Alternatives to Verification}
When verification is not cleared (e.g.,~the user wants to use song $17193$ from Table~\ref{tab:verification} as a reference song for parameter-level inference), a \textit{recommendation engine} can guide users towards compliant usage patterns that correspond with their underlying creative intentions. For example, a set of permitted inputs can be displayed to the user, allowing them to make compliant choices. Alternatively, for some use cases, a user could also be guided through a discovery journey to find existing music that meets their needs. These processes can be repeated through multi-turn interactions, until they converge to a point that satisfies the user's intent and the consent and licensing conditions of the underlying media.

\section{Generation}
\begin{figure}[t]
    \centering
    \includegraphics[width=0.95\linewidth]{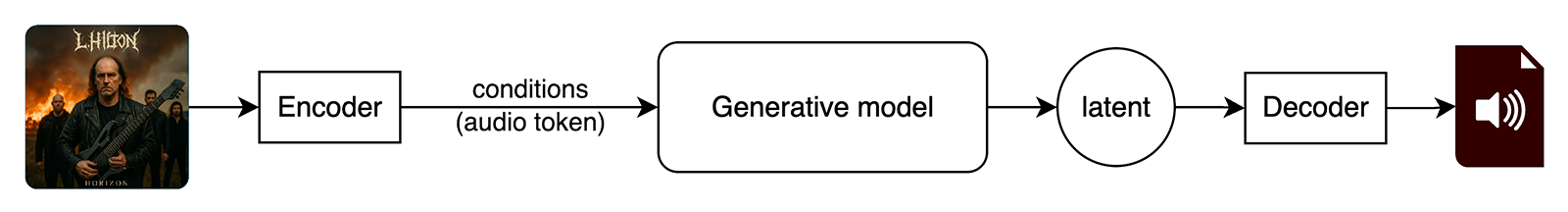}
    \caption{Pipeline to use a reference song to condition a generation.}
    \label{fig:gen1}
\end{figure}

\begin{figure}[t]
    \centering
    \includegraphics[width=1\linewidth]{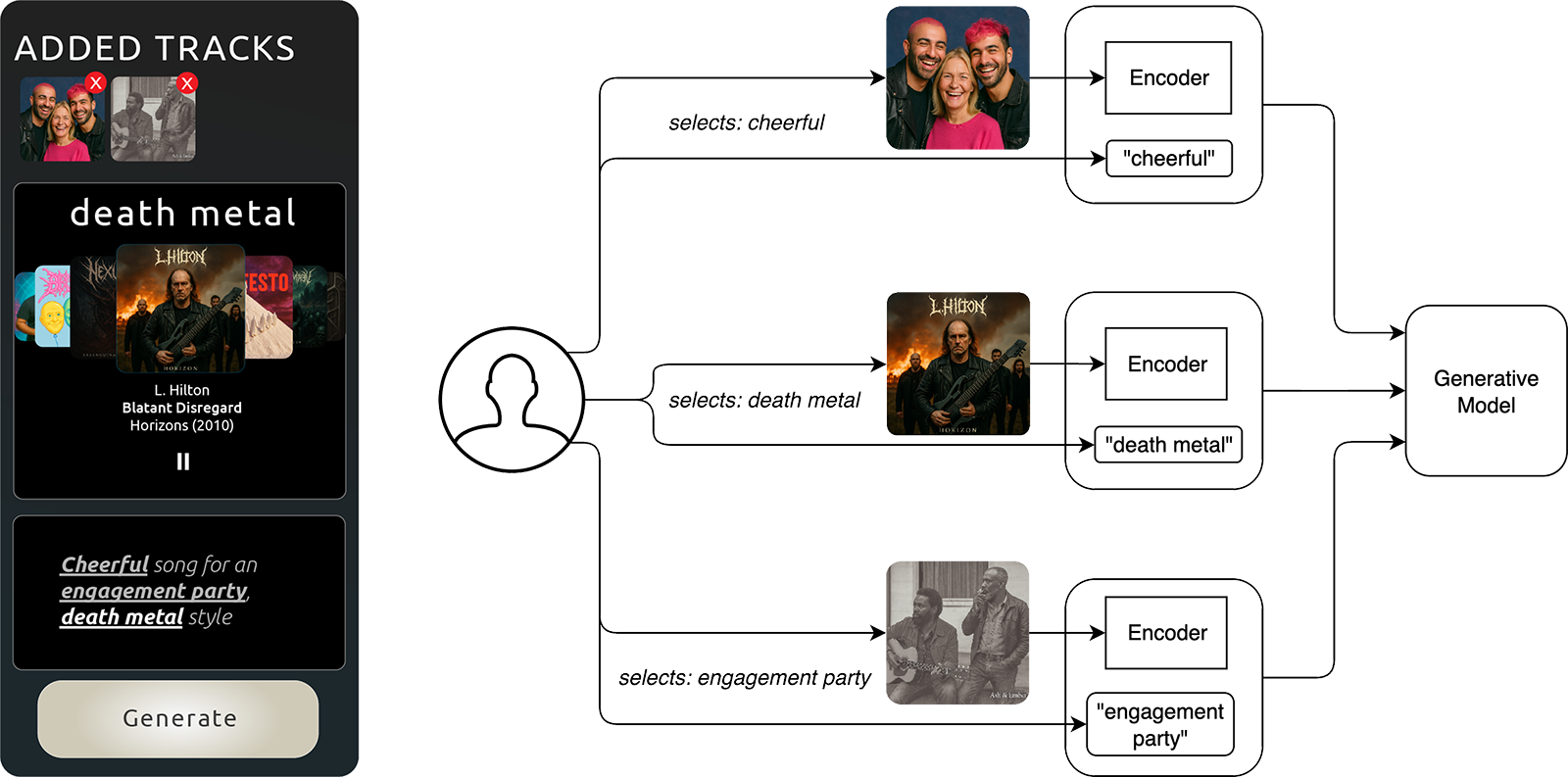}
    \caption{The generation stage needs to ensure that the functions of reference songs are respected. In the figure, the top song must influence the mood rather than the genre of the generation, which should instead be determined by the middle song.}
    \label{fig:flaggen}
\end{figure}

% Attribution-by-design essentially shifts the problem of determining influential songs from the training to the inference stage (from TTA to ITA), and develops a user interaction framework in which users actually define/choose such influential songs. 
%Reference audio can be seeded into a generative diffusion model to influence the latent trajectory using a Prompt-as-Sample generation approach.  \cite{park_understanding_2023} shows in the image domain that diffusion latents can be steered by embedding external signals, with early injections shaping coarse structure and later ones refining details; a similar strategy can be applied to audio to control both global form and local timbral features.
At the generation stage, the reference songs identified in the previous steps are passed through an encoder that generates audio tokens to be fed into the generative model (Fig.~\ref{fig:gen1}). This solution is straightforward when the user does not specify the \textit{function} of each of these reference songs, like in Fig.~\ref{fig:selection} (left and centre). In these cases, the user simply asks the system to generate music ``in the style of'' the reference material. Our framework, however, also supports more structured inputs, allowing users to assign different functions to specific songs. This is the case illustrated in Fig.~\ref{fig:selection} (right) and Fig.~\ref{fig:flaggen}. Here, each song has a specific function within the generation. In Fig.~\ref{fig:flaggen}, for instance, one reference song has been selected for its mood (cheerful), one for its genre (death metal), and one for the situation (an engagement party). In these cases, for each reference song, the audio embedding needs to be paired with a textual description (or just a flag) indicating its function in the generation (Fig.~\ref{fig:flaggen}). 

The situation becomes more complex in the case of parameter-level selection, like in Fig.~\ref{fig:selection} (right), when the user requested drawing the guitar style from one song and the melodic contour from another. Supporting such interactions requires the system to separate and recombine musical dimensions that are often deeply entangled in practice. Attributes such as melody, timbre, instrumentation, and affect are not orthogonal; a guitar texture, for instance, often conveys rhythmic, emotional, and even harmonic cues. This is also the case for categories like mood, genre, and situation, which are not independent but overlap with one another: a death metal track might not only convey genre signals, but also emotional weight, cultural valence, and situational affect. The imposition of categoric boundaries is reductive as it transforms continuous, entangled musical phenomena into discrete, manipulable units, a process which ~\cite{Morreale-2025} call \textit{grammatisation}. Nevertheless, assuming that these aspects can be partially disentangled and recombined is a reasonable and necessary abstraction for practical generation, provided that the model is trained to recognise and handle such relationships rather than treat them as perfectly separable categories.

To make this partial disentanglement and recombination possible, the model must be trained to account for such interactions. During training, the network should be exposed to structured examples that explicitly illustrate how different musical dimensions co-vary and can be recombined. One strategy is to include tuples that pair complementary roles drawn from different songs, annotated with functional flags (e.g., “melody from A,” “guitar texture from B”). Another strategy might be to provide multiple versions of the same composition, performed with different instruments (e.g., original, unplugged, or orchestral versions), so that the model encounters systematic variation in timbre and instrumentation while preserving the melodic and harmonic content (cf.~\citep{yesiler_audio-based_2021}). Such a design encourages the emergence of representations that are sensitive to relations among factors but flexible enough to recompose them meaningfully at inference time. In this way, disentanglement becomes not a rigid separation of dimensions, but a learned capacity to navigate their interplay.

Independent of the specific usage scenario, the underlying principle remains that the generative model anchors the synthesis of new material in the conditioning songs, ensuring that their influence remains central to the generated output and that attribution can be directly traced back to their creators. When disentanglement mechanisms are in place, this conditioning becomes more granular, allowing the model to selectively draw from specific aspects of each reference song (for example, its melody, instrumentation, or texture) rather than reproducing them as indivisible wholes. From a technical standpoint, this paradigm aligns well with existing works on retrieval-augmented generation~\citep{sheynin_knn-diffusion_2023,chen_re-imagen_2023,shalev-arkushin_imagerag_2025}, in the sense that the generative model is trained to extensively leverage and recycle the information contained in the conditioning data to generate new data. A further set of concepts that one could employ at inference time are the ones related to trajectory guidance and latent editing~\citep{ho_classifier-free_2021,park_understanding_2023,chen_exploring_2024}. Together, these components establish a generation process that is both transparent and attributable, while remaining sensitive to the complex entanglement of musical features that underpins creative recombination.

\section{Limitations and Future Work}

As TTA and ITA pose distinct epistemic and technical challenges, addressing both at once is unnecessary from a methodological perspective. In this paper, we focus on ITA, as TTA remains epistemically and technically uncertain. Should robust methods for TTA emerge, their results could be integrated into our framework, which is designed to evolve dynamically. We therefore regard TTA as a complementary line of inquiry that deserves dedicated investigation in future work. Two interim solutions for TTA could be envisioned. One solution is a flat-rate compensation scheme for artists whose catalogues have contributed to the training of the generative module. Another solution is to distribute royalties according to predictive attribution estimates derived from existing methods~\citep{deng_computational_2024,choi_large-scale_2025}. While inevitably uncertain, these methods might still provide a fairer basis for remuneration than a uniform flat rate.

Furthermore, a key challenge in the current \textit{mediated scenario} lies in the LMM's ability to retrieve relevant reference material from the inference dataset, which is constrained by both the scope and the diversity of its training data and the inference dataset. As a result, a system trained on and with access to narrow material at inference stage will struggle to interpret or retrieve relevant songs for user queries that reference rare, specific, or previously unseen audio scenarios. This limitation can lead to unsatisfactory user experiences when the desired reference material is either absent from the indexed catalogue or not well represented in the model’s embedding space. 

Finally, in addition to the generative model itself, generative systems in music typically incorporate several auxiliary modules, such as audio encoder/decoder structures~\citep{zeghidour_soundstream_2022,evans_fast_2024}, cross-modal alignment models~\citep{huang_mulan_2022,elizalde2022claplearningaudioconcepts}, or text tokenizers~\citep{lewis2019bart}. Like the core generative model, these modules are trained on large collections of human-generated audio, music, and text. As highlighted by \citet{morreale2025human}, such data are often repurposed from sources that were never intended for training generative models.  Yet, attribution discussions have so far focused almost exclusively on the output of the generative model, overlooking the fact that auxiliary modules, which are indispensable for the system’s operation, also rely on human work \citep{morreale2024unwitting}. As all these models can, in theory, be retrained, an ethical attribution-by-design model should not fall short of addressing this limitation and should offer some form of compensation to data creators for using their data to train these modules. 
% In this respect, the relevance of training data differs across submodules. Low-level codecs such as EnCodec \citep{défossez2022highfidelityneuralaudio} may have only a marginal impact on the specific characteristics of a generation, making attribution less critical in practice. By contrast, alignment models such as CLAP directly shape how user prompts are interpreted, since the quality of audio–text pairs in their training data determines what material the system is able to retrieve and how well it aligns with user requests. For such modules, extending attribution-by-design to cover the creators of caption–music data may be not only an ethical obligation but also a pragmatic business strategy. One could even imagine scenarios where captions are authored by the artists themselves, ensuring fair compensation while simultaneously improving the quality of text–music pairs and, consequently, the performance of the model. Notably, additional lump-sum compensation can be offered to artists who contribute textual descriptions about their songs to train auxiliary modules.

\section{Conclusions}
In this paper, we presented a paradigm shift for data attribution in generative systems, the innovative aspects and benefits of which are summarised as follows:
\begin{itemize}
    \item Training and inference datasets are handled separately. This separation has important consequences as it ontologically disentangles training-time from inference-time attribution.
    \item We specifically focused on inference time attribution, proposing a way to identify with certainty what specific inference data contributed to a given generation. Provenance of ``inspiration'' is not probabilistic, nor speculative. Rather, it is hard-coded into the architecture. Thus, our approach departs from mainstream research in attribution, which aims to estimate corroborative or predictive attribution scores. 
    \item Attribution-by-design is not simply an attribution method but is part of a sociotechnical architecture that enables hard-coding and automating all aspects related to the generation of an artwork: form collecting artists' preferences for how their catalogue is used by the various models embedded in the system to paving the way for automated royalty distribution.
    \item This approach has important benefits for the users, too: they can condition generations directly on existing songs of their liking and know for certain that part of their money directly goes to the referenced artist.
\end{itemize}

\section*{Ethical Disclosure}
We acknowledge the potentially disruptive impact of generative AI systems on the artistic industry. Our work specifically aims to redirect the efforts of AI researchers towards more ethical solutions that prioritise financial benefits for artists and rights holders, ensuring that creators are fairly compensated when their works contribute to AI-generated outputs. However, we also recognise that generative modelling may nevertheless inevitably displace certain artists and that not everyone will be in a position to receive compensation from a framework like the one we outline in this paper. To mitigate these risks, the implemented LMM might be designed in a way that is positively biased towards these artists. Furthermore, we suggest coupling attribution-by-design with complementary mechanisms such as public funds to support underrepresented creators, collective licensing schemes, or targeted subsidies for those whose works are less likely to be monetised through direct attribution. We also recognise that both training and operating such systems incur significant sustainability costs. We strongly advocate for the adoption of practices to offset the carbon footprint associated with these technologies. Finally, we are aware that the feasibility of implementing and deploying the proposed framework may be limited to companies and organisations with sufficient financial resources and access to a large music catalogue. We encourage further research and collaboration to make such systems more accessible and equitable across the industry.

\bibliographystyle{abbrvnat}
\bibliography{template}  

\end{document}